\documentclass[twocolumn,prl,aps,superscriptaddress,showpacs]{revtex4-1}
\usepackage[latin9]{inputenc}
\usepackage{amsmath}
\usepackage{amssymb}
\usepackage{graphicx}

\makeatletter

\DeclareTextSymbolDefault{\textquotedbl}{T1}

\usepackage{amsfonts}
\usepackage{color}
\usepackage{soul}

\makeatother

\begin{document}

\title{Deterministic generation of hybrid high-N00N states \\with Rydberg ions trapped in microwave cavities}

\author{Naeimeh Mohseni}
\email[]{n.mohseni@iasbs.ac.ir}
\affiliation{Department of Physics, Institute for Advanced Studies in Basic Sciences (IASBS), Iran}
\affiliation{Max-Planck-Institut f\"{u}r die Physik des Lichts, Staudtstrasse 2, 91058 Erlangen, Germany}

\author{Shahpoor Saeidian}
\email[]{saeidian@iasbs.ac.ir}
\affiliation{Department of Physics, Institute for Advanced Studies in Basic Sciences (IASBS), Iran}

\author{Jonathan P. Dowling}
\email[]{jdowling@phys.lsu.edu}
\affiliation{Hearne Institute for Theoretical Physics and Department of Physics and Astronomy, Louisiana State University,  Baton Rouge, LA 70803 USA}
\affiliation{CAS-Alibaba Quantum Computing Laboratory, USTC, Shanghai 201315, China
}
\affiliation{NYU-ECNU Institute of Physics at NYU Shanghai, Shanghai 200062, China}
\affiliation{National Institute of Information and Communications Technology,
Tokyo 184-8795, Japan
}

\author{Carlos Navarrete-Benlloch}
\email[]{derekkorg@gmail.com}
\affiliation{Max-Planck-Institut f\"{u}r die Physik des Lichts, Staudtstrasse 2, 91058 Erlangen, Germany}

\begin{abstract}

Trapped ions are among the most promising platforms for quantum technologies. They are at the heart of the most precise clocks and sensors developed to date, which exploit the quantum coherence of a single electronic or motional degree of freedom of an ion. However, future high-precision quantum metrology will require the use of entangled states of several degrees of freedom. Here we propose a protocol capable of generating high-N00N states where the entanglement is shared between the motion of a trapped ion and an electromagnetic cavity mode, a so-called `hybrid' configuration. We prove the feasibility of the proposal in a platform consisting of a trapped ion excited to its circular-Rydberg-state manifold, coupled to the modes of a high-Q microwave cavity. This compact hybrid architecture has the advantage that it can couple to signals of very different nature, which modify either the ion's motion or the cavity modes. Moreover, the exact same setup can be used right after the state-preparation phase to implement the interferometer required for quantum metrology.

\end{abstract}

\maketitle

\textbf{Introduction.} Trapped ions are at the forefront of quantum technological applications. They were the platform of choice for the first realistic quantum computing proposal \cite{Cirac95}, and are still part of the most promising architectures for such a long term goal \cite{Bermudez17,Monz16,Gaebler15}. More recently, they have been used to perform digital quantum simulations of relevance for condensed matter \cite{Debnath18,Jurcevic17,Lee16,Bohnet16,Clos16,Richerme14,Islam13,Islam11,Lanyon11,Kim10,Friedenauer08}, open systems \cite{Schindler13}, high-energy physics \cite{Martinez16,Muschik17}, quantum chemistry \cite{Hempel18}, and quantum optics \cite{Lv18}. But metrological applications are where trapped ions have traditionally shinned the brightest, owing to their robust quantum coherence, high controllability, and clean measurement protocols \cite{sinclair2014introduction,wineland2011quantum}. Indeed, the most precise clock built so far is based on the coherent transition between two internal states of a single trapped ion \cite{Huntemann16}.

However, moving forward in the field of quantum metrology will require exploiting more than just the quantum coherence of a single system. Indeed, it is by now well established that distributing entanglement among several systems can bring sensitivities all the way down to the ultimate Heisenberg limit \cite{dowling2008}. On this regard, N00N states of two oscillators and GHZ states of $N$ two-level systems are among the most promising entangled states. GHZ states have enjoyed a more successful experimental life, with states up to $N=14$ and $N=10$ generated with ion chains \cite{Monz11} and linear optics \cite{Wang16}, respectively. However, their use in quantum metrology requires the coherent manipulation of the large number of two-level systems, as well as their common coupling to the signal one wants to measure. In contrast, N00N states require the manipulation of just two harmonic modes (and only one of them has to couple to the signal), and are therefore more desirable in general. Unfortunately, high-N00N states have been traditionally more elusive. In the photonic case, the largest N00N state to date had $N=5$ \cite{Afek10}. In the case of ions, a big step forward has been recently achieved with the generation of an $N=9$ state of two motional modes \cite{Zhang16}.

If trapped ions are to come ahead also in this new `entangled metrological era', we will need to design further practical protocols for the generation of high-N00N states, either between ionic degrees of freedom, or between an ion and some other system, in what are dubbed `hybrid' configurations, which might lead to more flexible and versatile meters.

Here we show that Rydberg ions trapped in microwave cavities will open the door to the possibility of generating such states. These are novel platforms that have just taken its first steps \cite{Higgins17,Higgins17b}, and are set to combine the best of two well-established worlds: cavity quantum electrodynamics (QED), with microwave transitions between circular Rydberg states \cite{Raimond01,Haroche12}, and ions confined by radio-frequency traps \cite{Leibfried03,Wineland12}. In the former, GHz-range characteristic frequencies and high-Q microwave cavities allow access to the deep-strong coupling regime of light-matter interactions, while the latter is arguably among the most versatile quantum systems, allowing for the engineering of a large variety of effective interactions between motional, electronic, and photonic degrees of freedom. We exploit their combined outstanding properties to introduce an efficient and realistic protocol for the generation of hybrid high-N00N states of an ion and a cavity mode, in a compact architecture that can serve directly as the interferometer required for quantum metrology \cite{dowling2008}.

\begin{figure}[t]
\centering
\includegraphics[width=0.9\columnwidth]{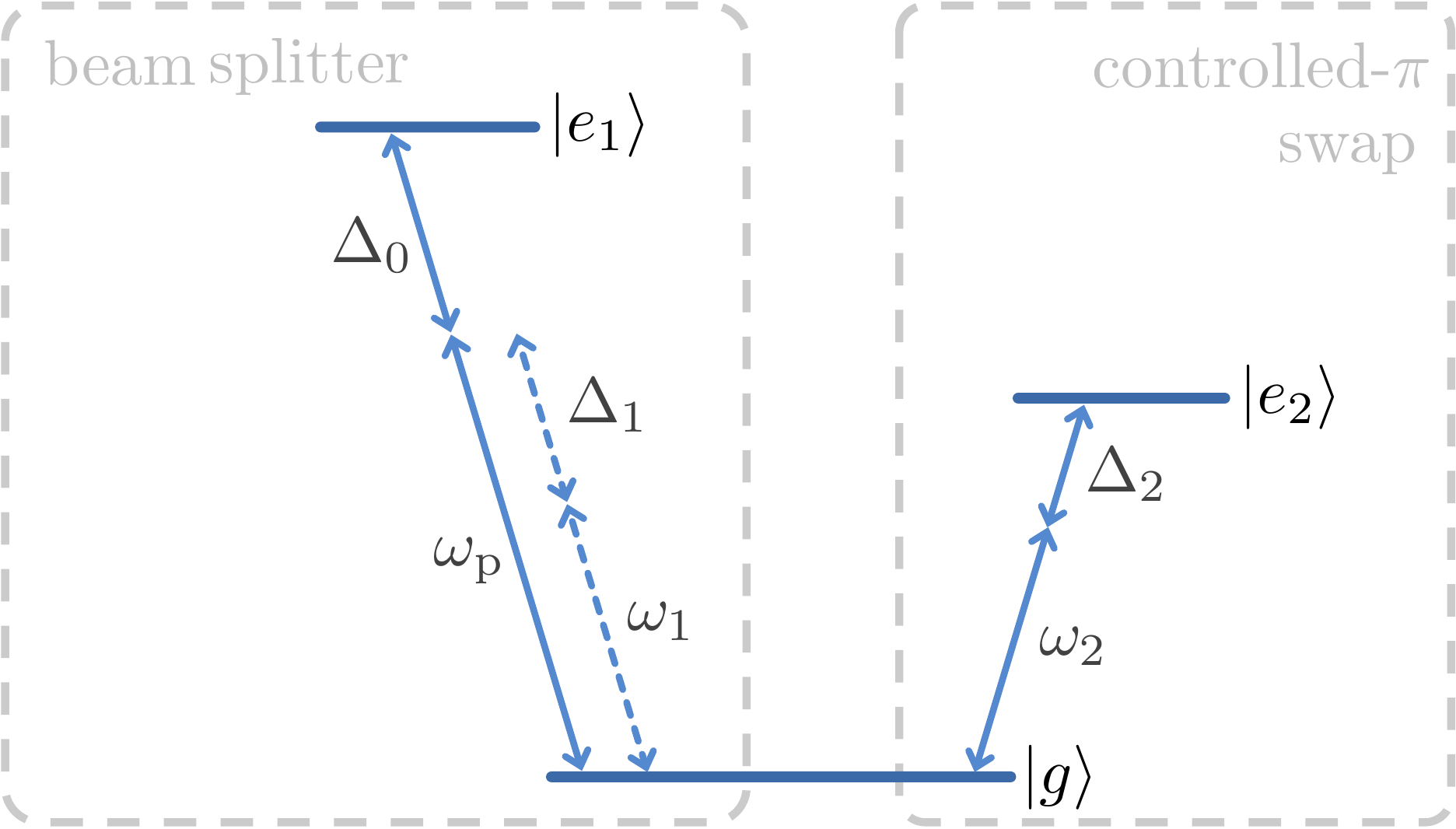}
\caption{Energy level scheme of the three-level ion interacting with the cavity modes. The first transition is used to implement a hybrid beam splitter when $|\Delta_0|\gg\Delta_1=\nu$ (together with more conditions detailed in the text). The second transition is used to implement a hybrid controlled-$\pi$ operation when $|\Delta_2|$ is large, or a swap operation (through a resonant Jaynes-Cummings interaction) when it is zero. Here we assume that the detunings $\Delta_{1,2}$ are tunable at real time, as explained in the text. \label{Fig:EnergyScheme}}
\end{figure}

Our protocol is inspired by the so-called ``magic'' beam splitter \cite{PhysRevA.64.063814}, which uses $N$-photon input states, beam splitters, controlled-$\pi$ phase gates, and on/off detectors as resources. Here we show that a trapped three-level ion interacting with two cavity modes provides all the required ingredients for the realistic implementation of a similar protocol, which we introduce in three steps. First, we propose an implementation of a hybrid beam splitter (HBS) between a cavity mode and a motional mode of the ion, with transmissivity and relative phase fully controllable via the amplitude of an external field and the interaction time. Hence, we call this a `temporal analog' of a HBS. We then show that a second internal transition can be used to generate strong hybrid cross-Kerr effective interactions between the cavity and the ion's motion, which implement the required controlled-phase gate. These operations are then combined with other standard ones and with the possibility of creating high-$N$ Fock states in microwave cavities \cite{Peaudecerf13,Zhou12,Sayrin11,Varcoe00} or in the ion's motion \cite{Kienzler17,Um16,BenKish03}, to show that three internal levels, together with a simple design of external drives and cavity interactions, suffice to implement a temporal analog of a magic beam splitter. After this, we assess in detail the feasibility of the proposal in foreseeable cavity-trapped Rydberg ions, estimating the effect of parameter fluctuations.

At the end we briefly comment on specific metrological applications, emphasizing that our hybrid N00N states provide a switchable photonic/mechanical architecture that is sensitive to signals of very different nature. Moreover, the hybrid Mach-Zehnder interferometer required for quantum enhanced metrology \cite{dowling2008} can be implemented directly with the same tools introduced for the N00N state generation.

\textbf{Temporal analog of a hybrid beam splitter.}
Let us consider an ion of mass $m$ cooled and confined by a one-dimensional harmonic potential \cite{Leibfried03} with trapping frequency $\nu$. We assume that the ion has been excited to a long-lived circular Rydberg state \cite{Raimond01}, and consider the transition between two of such internal states $|g\rangle$ and $|e_1\rangle$ with frequency difference $\omega_0$. The ion is inside a microwave Fabry-Perot cavity, and hence interacts with its standing-wave modes, of which we consider here a specific one with frequency $\omega_1$. The cavity field is pumped by a coherent microwave source at frequency $ \omega_\mathrm{p}$ (see the energy level scheme in Fig. \ref{Fig:EnergyScheme}). In a picture rotating at this frequency, the Hamiltonian reads \cite{Luo98}
\begin{eqnarray}\label{H_IonGen}
\hat{H}&=& \hbar \nu \hat{a}^{\dagger} \hat{a}+\hbar \Delta_0\hat{\sigma}_1^\dagger\hat{\sigma}_1+\hbar \Delta_1 \hat{b}^{\dagger}_1\hat{b}_1
\\
&& \hspace{3mm}-\hbar(\mathcal{E}^{*}\hat{b}_1+\mathcal{E}\hat{b}_1^{\dagger})+\hbar g(\hat{x}) (\hat{\sigma}_1^\dagger\hat{b}_1+\hat{\sigma}_1\hat{b}_1^{\dagger}), \nonumber
\end{eqnarray}
where $\hat{b}_1$ and $\hat{a}$ annihilate, respectively, cavity photons and motional quanta (phonons), $\hat{\sigma}_1=|g\rangle\langle e_1|$ is the lowering operator of the internal transition, $\mathcal{E}$ is the strength of the pump, which is detuned by $\Delta_j=\omega_j-\omega_\mathrm{p}$ from the corresponding frequency, $g(\hat{x})=\Omega \sin(\eta\hat{x}+\Phi)$, where $\Omega$ is the ion-cavity coupling strength (vacuum Rabi frequency), $\hat{x}=\hat{a}+\hat{a}^{\dagger}$, and $\Phi=\omega_1 x_0/c$, where $x_0$ is the position of the ion relative to a node of the standing wave. $\eta$ is the Lamb-Dicke parameter. Since we work with microwave modes, the bare $\eta$ (given by the ratio between the zero-point spatial fluctuations of the ion in the trap and the mode wavelength) is exceedingly small. However, it has been shown that $\eta$ is greatly enhanced in the presence of a magnetic field gradient \cite{Mintert01,Ospelkaus08,Johanning09,Ospelkaus11,Khromova12,Lake15,Piltze16,Weidt16,Wolk17}, capable of bringing it to the common regime $\eta\approx 0.1$ which we will assume in the following.

We consider the large-ionic-detuning limit ($|\Delta_0|\gg \Omega,|\Delta_1|,\nu$), where the internal levels can be adiabatically eliminated \cite{bhattacherjee2016optomechanical}. In  Ref. \cite{SupMat} we show that this leads to an effective Hamiltonian
\begin{equation} \label{H_OM} 
\hat{H}_\mathrm{OM}= \hbar \nu \hat{a}^{\dagger} \hat{a}\hspace{-0.2mm}+\hbar [\Delta_1-g_0(\hat{a}+\hat{a}^{\dagger})] \hat{b}_1^{\dagger}\hat{b}_1-\hbar(\mathcal{E}^{*}\hat{b}_1+\mathcal{E}\hat{b}_1^{\dagger}),
\end{equation}
where we have defined $g_0=\eta\Omega^{2}/2\Delta_0$, and have assumed that the ion is located in between a node and an anti-node ($\Phi=\pi/4$), which maximizes this effective coupling. This Hamiltonian is equivalent to that found in cavity quantum optomechanics \cite{aspelmeyer2014cavity}.

Including optical and motional damping at rates $\gamma$ and $\Gamma$, respectively, the master equation governing the evolution of the system can be written as
\begin{equation}\label{MasterEq}
\frac{d\hat{\rho}}{dt}=\bigg[\frac{\hat{H}_\mathrm{OM}}{i\hbar},\hat{\rho}\bigg]+\gamma\mathcal{D}_{b_1}[\hat{\rho}]+\Gamma\mathcal{D}_a[\hat{\rho}],
\end{equation} 
with dissipator $\mathcal{D}_J[\hat{\rho}]=2\hat{J}\hat{\rho}\hat{J}^\dagger-\hat{\rho}\hat{J}^\dagger\hat{J}-\hat{J}^\dagger\hat{J}\hat{\rho}$. As we show in Ref. \cite{SupMat}, the classical limit predicts a coherent state with amplitudes $\alpha$ and $\beta$ for the motional and optical modes, satisfying
\begin{subequations}
\begin{eqnarray}
\dot{\alpha}&=&-(\Gamma+i\nu)\alpha+ig_0|\beta|^2,
\\
\dot{\beta}&=&-[\gamma+i\Delta_1-ig_0(\alpha+\alpha^*)]\beta+i\mathcal{E}.
\end{eqnarray}
\end{subequations}
We will work under conditions \cite{SupMat} leading to a steady state $\bar{\alpha}=g_0|\bar{\beta}|^2/(\nu-i\Gamma)$ and $\bar{\beta}=\mathcal{E}/[\Delta_1-g_0(\bar{\alpha}+\bar{\alpha}^*)-i\gamma]$. Next, we consider small quantum fluctuations around it, by moving to a picture displaced to the classical solution and considering only terms in the master equation bilinear in annihilation and creation operators \cite{SupMat}. In this picture, the transformed state evolves according to Eq. (\ref{MasterEq}), but replacing the effective Hamiltonian by
\begin{equation}\label{H_lin}
\hat{H}_\mathrm{LIN}= \hbar\nu\hat{a}^\dagger\hat{a}+\hbar\Delta_1\hat{b}_1^\dagger\hat{b}_1-\hbar g_0 (\hat{a}+\hat{a}^\dagger)(\bar{\beta}^{*}\hat{b}_1+ \bar{\beta}\hat{b}_1^\dagger).
\end{equation} 
As we show in Ref. \cite{SupMat}, this `linearization' is valid provided that $|\beta|$ or $\nu/g_0$ are much larger than $\sqrt{N}$, where $N$ characterizes the photon number $\langle\hat{b}_1^\dagger\hat{b}_1\rangle$ in the displaced picture (which we anticipate matches the size of the N00N state).

Finally, choosing a detuning $\Delta_1=\nu$, and working in the $\nu\gg g_0|\bar{\beta}|$ regime, this Hamiltonian takes the form
\begin{equation}\label{H_BS}
\hat{H}_\mathrm{BS}= \hbar\nu(\hat{a}^\dagger\hat{a}+\hat{b}_1^\dagger\hat{b}_1)-\hbar g_0 (\bar{\beta}\hat{a}\hat{b}_1^\dagger+\bar{\beta}^{*}\hat{a}^\dagger\hat{b}_1),
\end{equation} 
within the rotating-wave approximation. The corresponding time evolution operator corresponds to a HBS operation whose mixing angle $\theta=g_0|\bar{\beta}|t$ and phase $\mathrm{arg}\{\bar{\beta}\}+\pi/2$ (assumed 0 in the following without loss of generalization) can be controlled via the interaction time and the pump amplitude $\mathcal{E}$. Note, however, that the performance is limited by the coherence time of the system, which we can estimate as $\gamma^{-1}$, typically much shorter than $\Gamma^{-1}$ \cite{Leibfried03}. Later we show that foreseeable Rydberg ions trapped in microwave cavities will allow for large enough coherence times leading to good fidelities for all the operations required in our protocol.

\textbf{Temporal analog of a hybrid controlled-phase gate}. We consider now the interaction between the ion and second cavity mode with frequency $\omega_2$, closer to resonance with a transition to a different excited state $|e_2\rangle$, but still detuned by $\Delta_2$ (see Fig. \ref{Fig:EnergyScheme}). The Hamiltonian takes the form (\ref{H_IonGen}), which assuming the ion to be located at the anti-node of the cavity mode ($\Phi=\pi/2$), leads to
\begin{equation}\label{H_IonAntinode}
\hat{H}=\hbar \nu \hat{a}^{\dagger} \hat{a}+\hbar \Delta_2 \hat{b}_2^{\dagger} \hat{b}_2+ \hbar \Omega\cos(\eta\hat{x})(\hat{\sigma}_2^\dagger\hat{b}_2+\hat{\sigma}_2\hat{b}_2^\dagger), 
\end{equation}
where in this case we are in a picture rotating at the frequency of the internal transition, and $\hat{b}_2$ and $\hat{\sigma}_2$ refer to the corresponding cavity mode and internal transition. In Ref. \cite{SupMat} we show that working in the $\nu\gg|\Delta_2|\gg\Omega$ regime, the adiabatic elimination of the internal levels leads to the effective hybrid cross-Kerr interaction \cite{semiao2005effective}
\begin{equation}\label{H_cKerr}
\hat{H}_\text{cK}=\hbar\nu\hat{a}^\dagger\hat{a} + \hbar\Delta_2\hat{b}_2^{\dagger}\hat{b}_2 - \hbar g_\text{cK}\hat{a}^\dagger\hat{a}\hat{b}_2^\dagger\hat{b}_2.
\end{equation}
where $g_\text{cK}=2\eta^2\Omega^2/\Delta_2$. In this case, the time-evolution operator is equivalent to a controlled-phase operation, where one mode feels a phase shift that depends on the number of photons of the other. The cross-phase shift $g_\text{cK}t$ can be controlled in this case through the interaction time. A $\pi$ shift requires $g_\text{cK}>\gamma$, which we prove feasible with Rydberg ions in microwave cavities. 

\begin{figure}[t]
\centering
\includegraphics[width=\columnwidth]{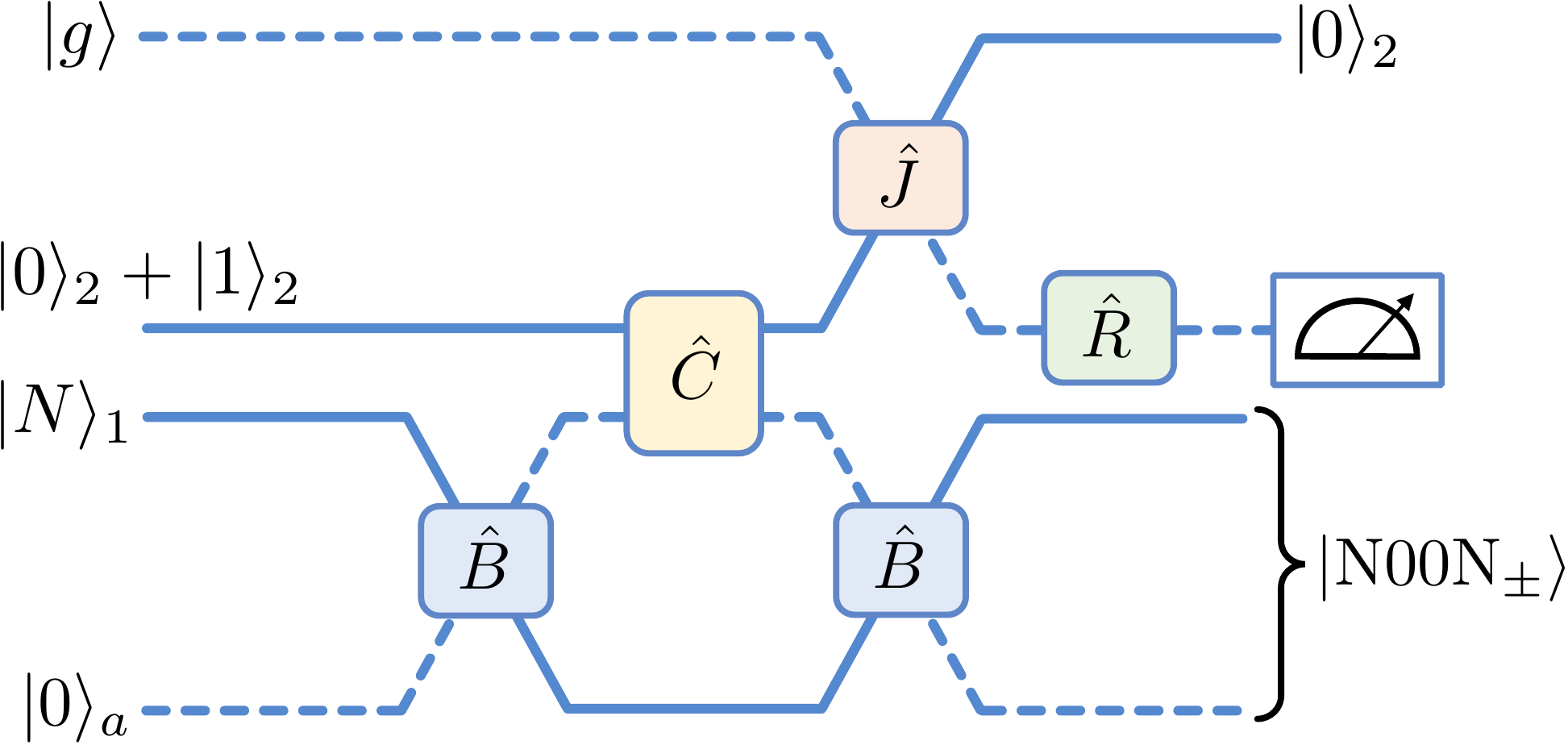}
\caption{Schematic representation of the protocol for the generation of N00N states. As detailed in the text, $a$ refers to the ion's motional mode, subindices $\{1,2\}$ to the cavity modes, $\hat{B}$ to a 50/50 beam splitter, $\hat{C}$ to a controlled-$\pi$, $\hat{J}$ to a swap, and $\hat{R}$ to a $\pi/4$ pulse. A final measurement of the internal state of the ion ($|g\rangle$ or $|e_2\rangle$) creates the desired N00N state. \label{Fig:NOONprotocol}}
\end{figure}
\textbf{High-N00N state generation protocol.} Our proposal is shown in Fig. \ref{Fig:NOONprotocol}, which is closely inspired the so-called magic beam splitter \cite{PhysRevA.64.063814}. In order to understand its principle of operation, let us first consider a situation without the controlled-$\pi$ operation ($\hat{C}$ yellow box in the figure), and follow the paths $1$ and $a$, which run along a Mach-Zehnder interferometer. With no more elements in the paths, the combination of the beam splitters (assumed the same and balanced) acts as a swap gate between the modes. Hence, starting with a Fock state with $N$ photons for definiteness (but the protocol works just as well starting with $N$ phonons instead), the state $|N\rangle_1|0\rangle_a$ turns into $|0\rangle_1|N\rangle_a$ (subindex $a$ refers to the motional Fock states, while subindices $1$ and $2$ refer to the corresponding cavity mode). The situation is radically different when a $\pi$ phase shift is performed on path $a$ in between the beam splitters, which completely cancels the effect of the latter. In such case, the input state remains unchanged. Hence, if one was able to engineer a balanced superposition of $0$ and $\pi$ phase shifts, the output state would turn into a superposition of $|N\rangle_1|0\rangle_a$ and $|0\rangle_1|N\rangle_a$, that is, a N00N state. This is exactly what is accomplished by the controlled-$\pi$ operation with mode $b_2$ (assumed in a superposition of $0$ and $1$ photons), together with the subsequent operations involving the second internal atomic transition. A final measurement revealing the state of the atom decides the relative phase between the $|N\rangle_1|0\rangle_a$ and $|0\rangle_1|N\rangle_a$ states forming the N00N state. In the remaining  part of this section we explain all these steps in detail, sticking to the operational (or gate) picture. In the next part we will then comment on the experimental requirements and their feasibility.

Recall that a balanced beam splitter acts as the unitary $\hat{B}=\exp[\pi(\hat{a}\hat{b}_1^\dagger-\hat{a}^\dagger\hat{b}_1)/4]$, which transforms
the operators as $\hat{B}\hat{a}\hat{B}^\dagger=(\hat{a}-\hat{b}_1)/\sqrt{2}$ and $\hat{B}\hat{b}_1\hat{B}^\dagger=(\hat{a}+\hat{b}_1)/\sqrt{2}$. Hence, applied to the initial state $|N\rangle_1|0\rangle_a(|0\rangle_2+|1\rangle_2)|g\rangle$ (we omit normalizations in the following to ease the notation), we obtain
\begin{equation}
(\hat{b}_1^\dagger+\hat{a}^\dagger)^N|0\rangle_1|0\rangle_a(|0\rangle_2+|1\rangle_2)|g\rangle,
\end{equation}
Next we apply the controlled-$\pi$ with unitary $\hat{C}=\exp(i\pi\hat{b}_2^\dagger\hat{b}_2\hat{a}^\dagger\hat{a})$, which turns the state into
\begin{equation}
(\hat{b}_1^\dagger+\hat{a}^\dagger)^N|0\rangle_1|0\rangle_a|0\rangle_2|g\rangle+(\hat{b}_1^\dagger-\hat{a}^\dagger)^N|0\rangle_1|0\rangle_a|1\rangle_2|g\rangle,
\end{equation}
where we have used $e^{i\pi\hat{a}^\dagger\hat{a}}\hat{a}e^{-i\pi\hat{a}^\dagger\hat{a}}=-\hat{a}$. A further beam splitter $\hat{B}$ then turns the state into
\begin{equation}
|0\rangle_1|N\rangle_a|0\rangle_2|g\rangle+|N\rangle_1|0\rangle_a|1\rangle_2|g\rangle.
\end{equation}
Finally, we apply two operations that involve the internal levels. First, an excitation-swap between the second cavity mode and the corresponding transition, with unitary $\hat{J}=\exp[-i\pi(\hat{b}_2^\dagger\hat{\sigma}_2+\hat{b}_2\hat{\sigma}_2^\dagger)/2]$, which effects the transformations $\hat{J}|0\rangle_2|g\rangle=|0\rangle_2|g\rangle$ and $\hat{J}|1\rangle_2|g\rangle=-i|0\rangle_2|e\rangle$. Then, we apply a $\pi/4$ pulse on the internal transition, with corresponding unitary $\hat{R}=\exp[\pi(\hat{\sigma}_2^\dagger-\hat{\sigma}_2)/4]$, and transformations $\hat{R}|g\rangle=|g\rangle+|e\rangle$ and $\hat{R}|e\rangle=|e\rangle-|g\rangle$. These turn the state into
\begin{equation}
|\text{N00N}_+\rangle|0\rangle_2|g\rangle+|\text{N00N}_-\rangle|0\rangle_2|e\rangle,
\end{equation}
where we have defined the hybrid N00N states $|\text{N00N}_\pm\rangle=|0\rangle_1|N\rangle_{a}\pm i|N\rangle_1|0\rangle_{a}$. It is then clear that a final measurement of the atomic state will project the ion and the first cavity mode into a $|\text{N00N}_\pm\rangle$ state depending on the outcome.

\textbf{Experimental considerations and feasibility.} Let us now comment on the experimental steps and corresponding requirements. We start with some general considerations, and at the end we will discuss specific parameters.

As we made obvious from the notation, the beam splitters $\hat{B}$ and controlled-$\pi$ $\hat{C}$ operations are implemented through cavity modes 1 and 2, respectively, following the methods presented in the first sections. The rest of operations are standard \cite{Raimond01}: the swap $\hat{J}$ is effected by letting a resonant Jaynes-Cummings (JC) interaction $\hbar\Omega (\hat{\sigma}_2^\dagger\hat{b}_2+\hat{\sigma}_2\hat{b}_2^{\dagger})$ run over a time $\pi/2\Omega$, while $\hat{R}$ is obtained by driving the second transition with a coherent microwave $\pi/4$ pulse. The crucial point is that, since the operations are applied sequentially, we need to be able to switch on and off the corresponding interactions at will. Here we suggest to do so by modifying the corresponding detunings in real time. In the case of the beam splitter interaction, this simply amounts to changing the pump frequency $\omega_\mathrm{p}$, which comes from an external source and is therefore easily tunable. In contrast, the real-time control of $\hat{C}$ and $\hat{J}$, operations involving the second internal transition, is more challenging because these do not involve external fields. However, it has been demonstrated that the frequency of the transition can be tuned in situ and fast through either the DC Stark or Zeeman shifts generated, respectively, by external electrostatic \cite{Raimond01,brune1996quantum} or magnetostatic \cite{Maurer04,Leibfried03} fields.

Another crucial piece is the preparation of the initial states. Specifically, in order to generate \textit{high}-N00N states, we need to initialize either the cavity mode 1 or the ion's motion in a Fock state with large $N$. Indeed, this has been achieved for both alternatives. In particular, Fock motional states up to $N=16$ were demonstrated in trapped ions more than twenty years ago \cite{Meekhof96} by exploiting the fact that the interaction between motion and internal states can be alternated between JC and anti-JC at will (see also \cite{Um16,Kienzler17,BenKish03} for more modern and elaborated experiments). In the case of microwave cavities, photonic Fock states up to $N=7$ have been stabilized via quantum feedback techniques \cite{Zhou12,Sayrin11,Peaudecerf13}. Finally, the preparation of cavity mode 2 in a superposition of 0 and 1 photons can be easily performed following techniques that have become standard in the field of cavity QED \cite{Raimond01}. For example, one can apply the inverse of the $\hat{R}\hat{J}$ sequence that we perform at the end of our protocol: starting from the atom in the ground state, a $\pi/4$ pulse is applied, followed by a swap of the internal excitation to the cavity field.

Note that the final measurement of the internal state can be performed following standard techniques in the field of trapped ions \cite{Leibfried03}. Hence, together with the previous discussion, this shows that all the pieces required for the implementation of the high-N00N protocol presented above are in principle available.

Let us now move on to the feasibility for concrete experimental parameters. The most demanding operation is the controlled-$\pi$, which requires the conditions $\nu\gg|\Delta_2|\gg\Omega$ together with $\eta^2\Omega^2/|\Delta_2|>\gamma$ in order to ensure that the coherence time is large enough to implement a $\pi$ shift. Taking as a reference cavity frequencies around $2\pi\times 10\mathrm{GHz}$, and taking into account that quality factors as large as $10^{10}$ are available in microwave cavities \cite{Varcoe00}, we obtain $\gamma=2\pi\times 10\mathrm{Hz}$. On the other hand, typical vacuum Rabi couplings for transitions between circular Rydberg states are around $\Omega=2\pi\times 50\mathrm{KHz}$. We then assume $|\Delta_2|=10\Omega$, and a trapping frequency $\nu=10|\Delta_2|=2\pi\times 5\mathrm{MHz}$, compatible with common traps \cite{Leibfried03,Kienzler17}. Putting all these estimates together, we obtain $\eta^2\Omega^2/|\Delta_2|=10\gamma$, as required.

The rest of operations are way less demanding. In particular, the effective beam splitter Hamiltonian requires the conditions $|\Delta_0|\gg\nu\gg g_0\mathrm{max}\{\sqrt{N},|\bar{\beta}|\}$, with $g_0|\bar{\beta}|\gg\gamma$ in order for coherence to be preserved during a sufficiently long time. Taking $|\Delta_0|=10\nu$, we obtain $g_0=2\pi\times 5\mathrm{Hz}$. The quantity $|\bar{\beta}|^2$, which provides the number of intracavity photons generated by the coherent microwave pump, is then bounded by $(\nu/g_0)^2=10^{12}\gg |\bar{\beta}|^2\gg(\gamma/g_0)^2=4$. Hence, choosing $|\bar{\beta}|^2$ between, e.g., 500 and 10000 (photon numbers easily generated with coherent pumps), we remain safely within the desired regime.

As for the swap operation, we simply need $\Omega\gg\mathrm{max}\{\gamma,\Gamma_e\}$, where $\Gamma_e$ refers to the spontaneous emission rate associated to the internal transition. For circular Rydberg states, the latter is typically on the tens of Hz \cite{Raimond01}, similarly to $\gamma$. Hence, we are deep into the required regime.

We have also analyzed the resilience of our proposal to parameter fluctuations. In particular, we have considered fluctuations in the beam splitter and controlled-$\pi$ parameters, finding analytic expressions for the fidelity, see Ref. \cite{SupMat} for details. As a figure of merit, we find fidelities above 90\% up to $N=40$ for a 1\% standard deviation in the parameters.

\textbf{Discussion and conclusions}. Our compact architecture offers unique opportunities from a metrological point of view. Its hybrid character makes it a versatile sensor, sensitive to signals that couple either to the electromagnetic field and the cavity or to the ion's motion. Moreover, right after generating the N00N state, the same setup can be used to implement the Mach-Zehnder interferometer required for metrology \cite{dowling2008}, which is essentially based on beam splitters.

In conclusion, we have shown that an architecture based on trapped ions excited to circular Rydberg states and coupled to the modes of a microwave cavity will be ideal for the compact implementation of a versatile quantum metrological system. Specifically, we have proposed a protocol for the generation of hybrid high-N00N states, showing its feasibility in near-future platforms. The same tools developed for the state-generation protocol (in particular the hybrid beam splitter) can be used to implement the interferometer required for quantum metrology, which will be sensitive to any signal that couples to either the cavity field or the motion of the ion.

\vspace{0.5cm}
\begin{acknowledgments}
This work was supported by the  Ministry of  Science Research and Technology of Iran and IASBS (Grant No, G 2018 IASBS 12648).
JPD would like to acknowledge support from the Air Force Office of Scientific Research, the Army Research Office, the Defense Advanced Projects Activity, the National Science Foundation, and the Northrop Grumman Corporation. NM would like to thank Marjan Fani for useful discussion. 
\end{acknowledgments}

\bibliography{Mohseni}

\begin{widetext}

\newpage

\begin{center}
\textbf{\Large{Supplemental material}}    
\end{center}

In this supplemental material we provide a detailed derivation of
the effective Hamiltonians introduced in the text. Specifically, we
first introduce the method of projectors, and use it to eliminate
the internal atomic levels, leading to optomechanical (2) and cross-Kerr
(8) Hamiltonians. Next, starting from the master equation (3),
we introduce the classical and linearized limits, eventually leading
to the linearized Hamiltonian (5). In the last part we provide further details about our analysis of parameter fluctuations.

\vspace{3mm}
\begin{center}
\textbf{\large{}I. Adiabatic elimination of the internal levels}{\large\par}
\end{center}

We start by providing a detailed elimination of the internal levels. We will proceed in the Schr\"odinger picture, using the method based on projection operators. Hence, we first introduce the general method, which we then particularize to the two relevant internal transitions.

\vspace{1mm}
\begin{center}
\textbf{I.A. General procedure: Projection operator method}
\end{center}

Let us introduce in general the method based on projection operators.
Consider a closed system evolving according to a Hamiltonian $\hat{H}$,
so that its state $|\psi(t)\rangle$ satisfies the Schr\"odinger equation
$\mathrm{i}\hbar\partial_{t}|\psi(t)\rangle=\hat{H}|\psi(t)\rangle$.
The idea of the method relies on the fact that we can divide the Hilbert
space into a relevant sector (whose effective dynamics we want to
describe) and an irrelevant one (whose dynamics is trivial, typically
because it stays unpopulated). We then define the projector operator
$\hat{P}=\hat{P}^{2}$, which projects onto the relevant subspace,
and its complement $\hat{Q}=1-\hat{P}$. Applying the latter onto
the Schr\"odinger equation, we get\begin{subequations}
\begin{align}
\mathrm{i}\hbar\partial_{t}\hat{Q}|\psi(t)\rangle & =\hat{Q}\hat{H}(\underset{1}{\underbrace{\hat{P}+\hat{Q}}})|\psi(t)\rangle=\hat{Q}\hat{H}\hat{Q}|\psi(t)\rangle+\hat{Q}\hat{H}\hat{P}|\psi(t)\rangle,\\
 & \Downarrow\nonumber \\
\hat{Q}|\psi(t)\rangle & =\frac{1}{\mathrm{i}\hbar}e^{\hat{Q}\hat{H}t/\mathrm{i}\hbar}\hat{Q}|\psi(0)\rangle+\int_{0}^{t}\frac{dt'}{\mathrm{i}\hbar}e^{\hat{Q}\hat{H}(t-t')/\mathrm{i}\hbar}\hat{Q}\hat{H}\hat{P}|\psi(t')\rangle.
\end{align}
\end{subequations}Naturally, we assume that the system is in the
relevant subspace initially, so that $\hat{Q}|\psi(0)\rangle=0$.
Hence, projecting the Schrödinger equation in the relevant subspace,
we then obtain
\begin{align}
\mathrm{i}\hbar\partial_{t}\hat{P}|\psi(t)\rangle & =\hat{P}\hat{H}\hat{P}|\psi(t)\rangle+\hat{P}\hat{H}\hat{Q}|\psi(t)\rangle=\hat{P}\hat{H}\hat{P}|\psi(t)\rangle+\int_{0}^{t}\frac{d\tau}{\mathrm{i}\hbar}\hat{P}\hat{H}e^{\hat{Q}\hat{H}\tau/\mathrm{i}\hbar}\hat{Q}\hat{H}\hat{P}|\psi(t-\tau)\rangle,\label{P_SchrEq}
\end{align}
where we have made the integration variable change $t'=t-\tau$.

With full generality, we can decompose the Hamiltonian as $\hat{H}=\hat{H}_{0}+\hat{H}_{1}$,
where $\hat{H}_{0}$ contains all the terms that do not connect the
relevant and irrelevant subspaces ($\hat{P}\hat{H}_{0}\hat{Q}=0=\hat{Q}\hat{H}_{0}\hat{P}$),
while $\hat{H}_{1}$ gathers the rest of the terms. Note that we can
even assume without loss of generality that $\hat{P}\hat{H}_{1}\hat{P}=0$,
that is, the `interaction' Hamiltonian does not connect states within
the relevant subspace. It is always possible to ensure such a property,
for if that's not the case, we just need to redefine $\hat{H}_{0}$
and $\hat{H}_{1}$ as $\hat{H}_{0}+\hat{P}\hat{H}_{1}\hat{P}$ and
$\hat{H}_{1}-\hat{P}\hat{H}_{1}\hat{P}$, respectively. Effective
theories are meaningful whenever one can treat $\hat{H}_{1}$ as a
perturbation with respect to $\hat{H}_{0}$. Hence, in the following
we consider only terms up to order two in $\hat{H}_{1}$ in (\ref{P_SchrEq}).
In order to do this, we use $|\psi(t-\tau)\rangle=e^{-\hat{H}\tau/\mathrm{i}\hbar}|\psi(t)\rangle$
and the property $\hat{P}\hat{H}_{0}\hat{Q}=0=\hat{Q}\hat{H}_{0}\hat{P}$,
which allows us to write (\ref{P_SchrEq}) as
\begin{align}
\mathrm{i}\hbar\partial_{t}\hat{P}|\psi(t)\rangle & =\hat{P}\hat{H}\hat{P}|\psi(t)\rangle+\int_{0}^{t}\frac{d\tau}{\mathrm{i}\hbar}\hat{P}\hat{H}_{1}e^{\hat{Q}\hat{H}\tau/\mathrm{i}\hbar}\hat{Q}\hat{H}_{1}\hat{P}e^{-\hat{H}\tau/\mathrm{i}\hbar}|\psi(t)\rangle\label{P_SchrEq-1}\\
 & =\left[\hat{P}\hat{H}\hat{P}+\int_{0}^{t}\frac{d\tau}{\mathrm{i}\hbar}\hat{P}\hat{H}_{1}e^{\hat{H}_{0}\tau/\mathrm{i}\hbar}\hat{H}_{1}e^{-\hat{H}_{0}\tau/\mathrm{i}\hbar}\hat{P}\right]\hat{P}|\psi(t)\rangle,\nonumber 
\end{align}
where in the last step we have made many simplifications. First, we
have neglected the $\hat{H}_{1}$ terms coming from the exponentials,
since the expression is already order two without counting them. We
have also made use of $[\hat{P},\hat{H}_{0}]=0=[\hat{Q},\hat{H}_{0}]$,
which follows directly from $0=\hat{P}\hat{H}_{0}\hat{Q}-\hat{Q}\hat{H}_{0}\hat{P}=\hat{P}\hat{H}_{0}(\hat{1}-\hat{P})-(\hat{1}-\hat{P})\hat{H}_{0}\hat{P}=[\hat{P},\hat{H}_{0}]$,
and we can use to prove the property

\begin{equation}
e^{\hat{Q}\hat{H}_{0}\tau/\mathrm{i}\hbar}=\sum_{k=0}^{\infty}\frac{1}{k!}\left(\frac{\tau}{\mathrm{i}\hbar}\right)^{k}\underset{k\text{ times}}{\underbrace{\hat{Q}\hat{H}_{0}\hat{Q}\hat{H}_{0}...\hat{Q}\hat{H}_{0}}}=\sum_{k=0}^{\infty}\frac{1}{k!}\left(\frac{\tau}{\mathrm{i}\hbar}\right)^{k}\underset{\hat{Q}}{\underbrace{\hat{Q}^{k}}}\hat{H}_{0}^{k}=\hat{Q}e^{\hat{H}_{0}\tau/\mathrm{i}\hbar}.
\end{equation}
Finally, we have used $\hat{P}\hat{H}_{1}\hat{Q}=\hat{P}\hat{H}_{1}(1-\hat{P})=\hat{P}\hat{H}_{1}$,
and similarly $\hat{Q}\hat{H}_{1}\hat{P}=\hat{H}_{1}\hat{P}$. The
term inside the brackets in (\ref{P_SchrEq-1}) can then be interpreted
as an effective Hamiltonian in the relevant subspace, which we can
write in the compact form
\begin{equation}
\hat{H}_{\text{eff}}(t)=\hat{P}\hat{H}_{0}\hat{P}+\int_{0}^{t}\frac{d\tau}{\mathrm{i}\hbar}\hat{P}\hat{H}_{1}\tilde{H}_{1}(\tau)\hat{P},\label{Heff2}
\end{equation}
with
\begin{equation}
\tilde{H}_{1}(\tau)=e^{\hat{H}_{0}\tau/\mathrm{i}\hbar}\hat{H}_{1}e^{-\hat{H}_{0}\tau/\mathrm{i}\hbar}.
\end{equation}
This provides the final expression we will work with.

Note that (\ref{Heff2}) is not Hermitian, which seems to be at odds
with the fact that we interpret it as an effective Hamiltonian. In addition,
(\ref{Heff2}) is time dependent, even if the original Hamiltonian
was time independent. However, in many situations it indeed occurs
that (\ref{Heff2}) becomes approximately Hermitian and time independent
under the same physical conditions that allow us to split the Hilbert
space into relevant and irrelevant subspaces. We will see this in
the examples that we treat next.

\begin{center}
\textbf{I.B. Elimination of $|e_{1}\rangle$: Effective optomechanical interaction}
\end{center}

Let us apply the general framework presented above to the elimination
of the first transition of the ion presented in the main text, described
by the Hamiltonian (1) that we reproduce here for convenience
\begin{eqnarray}
\hat{H} & = & \hbar\nu\hat{a}^{\dagger}\hat{a}+\hbar\Delta_{0}\hat{\sigma}_{1}^{\dagger}\hat{\sigma}_{1}+\hbar\Delta_{1}\hat{b}_{1}^{\dagger}\hat{b}_{1}-\hbar(\mathcal{E}^{*}\hat{b}_{1}+\mathcal{E}\hat{b}_{1}^{\dagger})+\hbar g(\hat{x})(\hat{\sigma}_{1}^{\dagger}\hat{b}_{1}+\hat{\sigma}_{1}\hat{b}_{1}^{\dagger}),\label{H_IonGen_SupMat}
\end{eqnarray}
with $g(\hat{x})=\Omega \sin(\eta\hat{x}+\Phi)$. Assuming that $|\Delta_{0}|\gg\Omega,|\Delta_{1}|,\nu,|\mathcal{E}|$,
with the ion starting in the ground state $|g\rangle$, we expect
the excited state $|e_{1}\rangle$ to remain unpopulated. Hence, the
Hilbert space is naturally devided into a relevant one described by
the projector $\hat{P}=|g\rangle\langle g|$ and an irrelevant one
with projector $\hat{Q}=1-|g\rangle\langle g|=|e_{1}\rangle\langle e_{1}|$.
Similarly, using the notation of the previous section, the Hamiltonian
is naturally split into\begin{subequations}
\begin{align}
\hat{H}_{0} & =\hbar\nu\hat{a}^{\dagger}\hat{a}+\hbar\Delta_{0}\hat{\sigma}_{1}^{\dagger}\hat{\sigma}_{1}+\hbar\Delta_{1}\hat{c}_{1}^{\dagger}\hat{c}_{1},\\
\hat{H}_{1} & =\hbar g(\hat{x})\left[\hat{\sigma}_{1}^{\dagger}\left(\hat{c}_{1}+\frac{\mathcal{E}}{\Delta_{1}}\right)+\hat{\sigma}_{1}\left(\hat{c}_{1}^{\dagger}+\frac{\mathcal{E}^{*}}{\Delta_{1}}\right)\right],
\end{align}
\end{subequations}
where, for convenience, we have defined the displaced photonic operator
$\hat{c}_{1}=\hat{b}_{1}-\mathcal{E}/\Delta_{1}$ (hence this Hamiltonian
defers from the previous one by a constant shift $|\mathcal{E}|^{2}/\Delta_{1}$,
irrelevant for the system dynamics). Taking into account that\begin{subequations}
\begin{align}
e^{\hat{H}_{0}\tau/\mathrm{i}\hbar}\hat{\sigma}_{1}e^{-\hat{H}_{0}\tau/\mathrm{i}\hbar} & =e^{i\Delta_{0}\tau}\hat{\sigma}_{1},\\
e^{\hat{H}_{0}\tau/\mathrm{i}\hbar}\hat{a}e^{-\hat{H}_{0}\tau/\mathrm{i}\hbar} & =e^{i\nu\tau}\hat{a},\\
e^{\hat{H}_{0}\tau/\mathrm{i}\hbar}\hat{c}_{1}e^{-\hat{H}_{0}\tau/\mathrm{i}\hbar} & =e^{i\Delta_{1}\tau}\hat{c}_{1},
\end{align}
\end{subequations}we then have
\begin{equation}
\tilde{H}_{1}(\tau)=e^{\hat{H}_{0}\tau/\mathrm{i}\hbar}\hat{H}_{1}e^{-\hat{H}_{0}\tau/\mathrm{i}\hbar}=\hbar g[\tilde{x}(\tau)]e^{-i\Delta_{0}\tau}\hat{\sigma}_{1}^{\dagger}\left(e^{i\Delta_{1}\tau}\hat{c}_{1}+\frac{\mathcal{E}}{\Delta_{1}}\right)+\mathrm{H.c.},	
\end{equation}
with
\begin{equation}
\tilde{x}(\tau)=e^{i\nu\tau}\hat{a}+e^{-i\nu\tau}\hat{a}^{\dagger}.	
\end{equation}
The second order term of the effective Hamiltonian (\ref{Heff2})
then reads
\begin{align}\label{Heff1t}
\int_{0}^{t}\frac{d\tau}{i\hbar}\hat{P}\hat{H}_{1}\tilde{H}_{1}(\tau)\hat{P} & =-i\hbar g(\hat{x})\left(\hat{c}_{1}^{\dagger}+\frac{\mathcal{E}^{*}}{\Delta_{1}}\right)\int_{0}^{t}d\tau g[\tilde{x}(\tau)]e^{-i\Delta_{0}\tau}\left(e^{i\Delta_{1}\tau}\hat{c}_{1}+\frac{\mathcal{E}}{\Delta_{1}}\right)|g\rangle\langle g|\\
&\approx -i\hbar g^{2}(\hat{x})\left(\hat{c}_{1}^{\dagger}+\frac{\mathcal{E}^{*}}{\Delta_{1}}\right)\left(\hat{c}_{1}+\frac{\mathcal{E}}{\Delta_{1}}\right)\int_{0}^{t}d\tau e^{-i\Delta_{0}\tau}|g\rangle\langle g|\nonumber \\
&= \hbar\frac{g^{2}(\hat{x})}{\Delta_{0}}\hat{b}_{1}^{\dagger}\hat{b}_{1}\left(e^{-i\Delta_{0}t}-1\right)|g\rangle\langle g|,\nonumber 
\end{align}
where we have performed an intermediate approximation neglecting the
oscillations at frequencies $\nu$ and $|\Delta_{1}|$ in the time integral,
compatible with the fact that the oscillations at $|\Delta_{0}|$ are
much faster (this step is not critical, that is, we can perform the
required time integrations including all time scales, but it simplifies
the derivation enormously, and leads to the same final result in the considered regime). The
time dependence of (\ref{Heff1t}) can be neglected within a rotating-wave approximation
as long as $|\Delta_{0}|\gg\Omega$.

Next we use $\eta\ll1$ to expand the coupling term as $g^{2}(\hat{x})=\Omega^{2}\sin^{2}(\eta\hat{x}+\Phi)\approx\Omega^{2}\left[\sin^{2}(\Phi)+\eta\hat{x}\sin(2\Phi)/2\right]$.
Choosing $\Phi=\pi/4$ as mentioned in the main text, we then obtain
the final effective Hamiltonian
\begin{equation}
\hat{H}_{\text{eff}}=\hbar\nu\hat{a}^{\dagger}\hat{a}+\hbar\left(\Delta_{1}-\frac{\Omega^{2}+\eta\Omega^{2}\hat{x}}{2\Delta_{0}}\right)\hat{b}_{1}^{\dagger}\hat{b}_{1}-\hbar(\mathcal{E}^{*}\hat{b}_{1}+\mathcal{E}\hat{b}_{1}^{\dagger}),
\end{equation}
which matches the optomechanical Hamiltonian (2) introduced in the main
text. Note that there we made the simplification $|\Delta_{1}|\gg\Omega^{2}/2|\Delta_{0}|$,
which is usually very well satisfied.

\begin{center}
\textbf{I.C. Elimination of $|e_{2}\rangle$: Effective cross-Kerr interaction}
\end{center}

We apply now the method to the second transition of the ion. The corresponding
Hamiltonian is described by (7), that is,
\begin{eqnarray}
\hat{H} & = & \hbar\nu\hat{a}^{\dagger}\hat{a}+\hbar\Delta_{2}\hat{b}_{2}^{\dagger}\hat{b}_{2}+\hbar\Omega\cos(\eta\hat{x})(\hat{\sigma}_{2}^{\dagger}\hat{b}_{2}+\hat{\sigma}_{2}\hat{b}_{2}^{\dagger}).\label{H_Ion2_SupMat}
\end{eqnarray}
In this case we assume that $\nu\gg|\Delta_{2}|\gg\Omega$.
Hence, for an ion starting in the ground state $|g\rangle$, we expect
again the excited state $|e_{2}\rangle$ to remain unpopulated. Therefore,
the projector onto the relevant subspace reads again $\hat{P}=|g\rangle\langle g|$,
so that $\hat{Q}=|e_{2}\rangle\langle e_{2}|$. We now split the Hamiltonian
into\begin{subequations}
\begin{align}
\hat{H}_{0} & =\hbar\nu\hat{a}^{\dagger}\hat{a}+\hbar\Delta_{2}\hat{b}_{2}^{\dagger}\hat{b}_{2},\\
\hat{H}_{1} & =\hbar\Omega\cos(\eta\hat{x})(\hat{\sigma}_{2}^{\dagger}\hat{b}_{2}+\hat{\sigma}_{2}\hat{b}_{2}^{\dagger}).
\end{align}
\end{subequations}Taking into account that\begin{subequations}
\begin{align}
e^{\hat{H}_{0}\tau/\mathrm{i}\hbar}\hat{\sigma}_{2}e^{-\hat{H}_{0}\tau/\mathrm{i}\hbar} & =\hat{\sigma}_{2},\\
e^{\hat{H}_{0}\tau/\mathrm{i}\hbar}\hat{a}e^{-\hat{H}_{0}\tau/\mathrm{i}\hbar} & =e^{i\nu\tau}\hat{a},\\
e^{\hat{H}_{0}\tau/\mathrm{i}\hbar}\hat{b}_{2}e^{-\hat{H}_{0}\tau/\mathrm{i}\hbar} & =e^{i\Delta_{2}\tau}\hat{b}_{2},
\end{align}
\end{subequations}we then have
\begin{equation}
\tilde{H}_{1}(\tau)=e^{\hat{H}_{0}\tau/\mathrm{i}\hbar}\hat{H}_{1}e^{-\hat{H}_{0}\tau/\mathrm{i}\hbar}=\hbar\Omega\cos[\eta\tilde{x}(\tau)]e^{i\Delta_{2}\tau}\hat{\sigma}_{2}^{\dagger}\hat{b}_{2}+\mathrm{H.c.}
\end{equation}
Before proceeding, it is now convenient to use the $\eta\ll1$ expansion
\begin{equation}
\cos[\eta\tilde{x}(\tau)]\approx1-\eta^{2}\tilde{x}^{2}(\tau)/2\approx1-\eta^{2}\hat{a}^{\dagger}\hat{a}-\eta^{2}\left(e^{2i\nu\tau}\hat{a}^{2}+e^{-2i\nu\tau}\hat{a}^{\dagger2}\right)/2,
\end{equation}
so that the second order term of the effective Hamiltonian (\ref{Heff2})
can be written as
\begin{align}
\int_{0}^{t}\frac{d\tau}{i\hbar}\hat{P}\hat{H}_{1}\tilde{H}_{1}(\tau)\hat{P}= & -i\hbar\Omega^{2}\cos(\eta\hat{x})\hat{b}_{2}^{\dagger}\hat{b}_{2}\int_{0}^{t}d\tau\cos[\eta\tilde{x}(\tau)]e^{i\Delta_{2}\tau}|g\rangle\langle g|\\
\approx & \hbar\Omega^{2}\cos(\eta\hat{x})\hat{b}_{2}^{\dagger}\hat{b}_{2}\left[(1-\eta^{2}\hat{a}^{\dagger}\hat{a})\frac{1-e^{i\Delta_{2}\tau}}{\Delta_{2}}-\frac{\eta^{2}}{2}\frac{1-e^{2i\nu t}}{\Delta_{2}+2\nu}\hat{a}^{2}-\frac{\eta^{2}}{2}\frac{1-e^{2i\nu t}}{\Delta_{2}-2\nu}\hat{a}^{\dagger2}\right]|g\rangle\langle g|\nonumber \\
\approx & \hbar\frac{\Omega^{2}}{\Delta_{2}}\hat{b}_{2}^{\dagger}\hat{b}_{2}(1-\eta^{2}\hat{a}^{\dagger}\hat{a})^{2}|g\rangle\langle g|,\nonumber 
\end{align}
where in the last approximation we have made use of the regime $\nu\gg|\Delta_{2}|\gg\Omega$,
which allows us to neglect all terms except the one presented at the
end. Combining this with the zeroth order term, and keeping terms up
to second order in $\eta$ we obtain the effective Hamiltonian
\begin{equation}
\hat{H}_{\text{eff}}=\hbar\nu\hat{a}^{\dagger}\hat{a}+\hbar\Delta_{2}\left(1+\frac{\Omega^{2}}{\Delta_{2}^{2}}\right)\hat{b}_{1}^{\dagger}\hat{b}_{1}-\hbar\frac{2\eta^{2}\Omega^{2}}{\Delta_{2}}\hat{b}_{2}^{\dagger}\hat{b}_{2}\hat{a}^{\dagger}\hat{a},
\end{equation}
which matches the cross-Kerr Hamiltonian (8) introduced in the main text, once $\Omega^{2}/\Delta_{2}^{2}$ is neglected in the parenthesis.

\begin{center}
\textbf{\large{}II. Linearization of the optomechanical interaction}{\large\par}
\end{center}

In this section we explain in detail the process of linearizing the
master equation (3), which we reproduce here for convenience
\begin{equation}
\frac{d\hat{\rho}}{dt}=\bigg[\frac{\hat{H}_{\text{OM}}}{i\hbar},\hat{\rho}\bigg]+\gamma\mathcal{D}_{b_{1}}[\hat{\rho}]+\Gamma\mathcal{D}_{a}[\hat{\rho}],\label{MasterEq_SupMat}
\end{equation}
with
\begin{subequations}
\begin{eqnarray}
\hat{H}_{\text{OM}}&=&\hbar\nu\hat{a}^{\dagger}\hat{a}+\hbar[\Delta_{1}-g_{0}(\hat{a}+\hat{a}^{\dagger})]\hat{b}_{1}^{\dagger}\hat{b}_{1}-\hbar\mathcal{E}^{*}\hat{b}_{1}-\hbar\mathcal{E}\hat{b}_{1}^{\dagger},
\\
\mathcal{D}_J[\hat{\rho}]&=&2\hat{J}\hat{\rho}\hat{J}^\dagger-\hat{J}^\dagger\hat{J}\hat{\rho}-\hat{\rho}\hat{J}^\dagger\hat{J}.
\end{eqnarray}
\end{subequations}

\vspace{3mm}
\begin{center}
\textbf{II.A. The classical limit}
\end{center}

Linearization consists in considering small quantum fluctuations around
the classical state of the system. Hence, we first consider here the
classical limit, which in this case is obtained by assuming that the
state is a product of coherent states for both modes: $|\alpha\rangle\otimes|\beta\rangle$,
where $\hat{a}|\alpha\rangle=\alpha|\alpha\rangle$ and $\hat{b}_{1}|\beta\rangle=\beta|\beta\rangle$.
The master equation can then be turned into an evolution equation
for the coherent amplitudes $\alpha(t)$ and $\beta(t)$. Let us find
such equation.

In order to do this, it is convenient to first note that the evolution equation of the expectation
value of any operator $\hat{A}$ can be written as
\begin{align}
\frac{d\langle\hat{A}\rangle}{dt} & =\mathrm{tr}\left\{ \hat{A}\frac{d\hat{\rho}}{dt}\right\} =\left\langle \left[\hat{A},\frac{\hat{H}_{\text{OM}}}{i\hbar}\right]\right\rangle +\gamma\left\langle [\hat{b}_{1}^{\dagger},\hat{A}]\hat{b}_{1}+\hat{b}_{1}^{\dagger}[\hat{A},\hat{b}_{1}]\right\rangle +\Gamma\left\langle [\hat{a}^{\dagger},\hat{A}]\hat{a}+\hat{a}^{\dagger}[\hat{A},\hat{a}]\right\rangle .
\end{align}
Applying this expression to the operators $\hat{a}$ and $\hat{b}_{1}$,
and using the fact that coherent states are their eigenstates, we
easily find the evolution equations\begin{subequations} \label{ClassicalEvoEqs}
\begin{eqnarray}
\dot{\alpha} & = & -(\Gamma+i\nu)\alpha+ig_{0}|\beta|^{2},\\
\dot{\beta} & = & -[\gamma+i\Delta_{1}-ig_{0}(\alpha+\alpha^{*})]\beta+i\mathcal{E}.
\end{eqnarray}
\end{subequations}These evolution equations possess stationary states ($\dot{\alpha}=0=\dot{\beta}$)
defined by\begin{subequations}\label{ClassicalSteady} 
\begin{align}
\bar{\alpha} & =g_{0}|\bar{\beta}|^{2}/(\nu-i\Gamma)\approx g_{0}|\bar{\beta}|^{2}/\nu,\\
\bar{\beta} & =\mathcal{E}/[\Delta_{1}-g_{0}(\bar{\alpha}+\bar{\alpha}^{*})-i\gamma]\approx\mathcal{E}/\nu,
\end{align}
\end{subequations} where in the last step we have made use of the
regime $\Delta_{1}=\nu\gg\mathrm{max}\{g_{0}|\bar{\beta}|,\gamma,\Gamma\}$
that we showed in the main text is required for an appropriate beam splitter operation. This solution must be stable
against perturbations in order for linearization to work. Writing
$\alpha(t)=\bar{\alpha}+\delta\alpha(t)$ and $\beta(t)=\bar{\beta}+\delta\beta(t)$
in (\ref{ClassicalEvoEqs}), and keeping terms to first order in the
fluctuations, we get
\begin{align}
\frac{d}{dt}\left(\begin{array}{c}
\delta\alpha\\
\delta\alpha^{*}\\
\delta\beta\\
\delta\beta^{*}
\end{array}\right) & =\underbrace{\left(\begin{array}{cccc}
-\Gamma-i\nu & 0 & ig_{0}\bar{\beta}^{*} & ig_{0}\bar{\beta}\\
0 & -\Gamma+i\nu & -ig_{0}\bar{\beta}^{*} & -ig_{0}\bar{\beta}\\
ig_{0}\bar{\beta} & ig_{0}\bar{\beta} & -\gamma-i\Delta_{1}+2ig_{0}\text{Re}\{\bar{\alpha}\} & 0\\
-ig_{0}\bar{\beta}^{*} & -ig_{0}\bar{\beta}^{*} & 0 & -\gamma+i\Delta_{1}-2ig_{0}\text{Re}\{\bar{\alpha}\}
\end{array}\right)}_{\mathcal{M}}\left(\begin{array}{c}
\delta\alpha\\
\delta\alpha^{*}\\
\delta\beta\\
\delta\beta^{*}
\end{array}\right)\\
 & \approx \left(\begin{array}{cccc}
-\Gamma-i\nu & 0 & ig_{0}\bar{\beta}^{*} & ig_{0}\bar{\beta}\\
0 & -\Gamma+i\nu & -ig_{0}\bar{\beta}^{*} & -ig_{0}\bar{\beta}\\
ig_{0}\bar{\beta} & ig_{0}\bar{\beta} & -\gamma-i\nu & 0\\
-ig_{0}\bar{\beta}^{*} & -ig_{0}\bar{\beta}^{*} & 0 & -\gamma+i\nu
\end{array}\right)\left(\begin{array}{c}
\delta\alpha\\
\delta\alpha^{*}\\
\delta\beta\\
\delta\beta^{*}
\end{array}\right),\nonumber 
\end{align}
where in the last step we have made use of the regime $\Delta_{1}=\nu\gg g_{0}|\bar{\beta}|$
and (\ref{ClassicalSteady}). The solution will be stable whenever the fluctuations decay towards
zero, which in turn happens only if the eigenvalues of $\mathcal{M}$, known as linear stability
matrix, have all negative real part. This is clearly the case for
$g_{0}=0$. On the other hand, the terms proportional to $g_{0}|\bar{\beta}|\ll\nu$
are just a small perturbation which is readily shown to not be able
to make the system unstable. Hence, in conclusion, under our operating
conditions it is ensured that the stationary solution (\ref{ClassicalSteady})
is stable.

\vspace{3mm}
\begin{center}
\textbf{II.B. Linearization of quantum fluctuations}
\end{center}

In order to introduce the linearized approximation for quantum fluctuations,
we move to a picture displaced to the classical steady state presented
above. Defining the displacement $\hat{D}(\bar{\alpha},\bar{\beta})=\exp(\bar{\alpha}\hat{a}+\bar{\beta}\hat{b}_1-\mathrm{H.c.})$
which transforms the bosonic operators as $\hat{D}^{\dagger}\hat{a}\hat{D}=\hat{a}+\bar{\alpha}$
and $\hat{D}^{\dagger}\hat{b}_{1}\hat{D}=\hat{b}_{1}+\bar{\beta}$,
the transformed state $\tilde{\rho}=\hat{D}^{\dagger}\hat{\rho}\hat{D}$
is easily shown to evolve according to the master equation
\begin{equation}
\frac{d\tilde{\rho}}{dt}=\bigg[\frac{\tilde{H}}{i\hbar},\tilde{\rho}\bigg]+\gamma\mathcal{D}_{b_{1}}[\tilde{\rho}]+\Gamma\mathcal{D}_{a}[\tilde{\rho}],\label{MasterEq_SupMat_Linearization}
\end{equation}
with
\begin{equation}
\tilde{H}=\hbar\nu\hat{a}^{\dagger}\hat{a}+\hbar\tilde{\Delta}_{1}\hat{b}_{1}^{\dagger}\hat{b}_{1}-\hbar g_{0}(\hat{a}+\hat{a}^{\dagger})(\bar{\beta}\hat{b}_{1}^{\dagger}+\bar{\beta}^{*}\hat{b}_{1}+\hat{b}_{1}^{\dagger}\hat{b}_{1}),
\end{equation}
with $\tilde{\Delta}_{1}=\Delta_{1}-g_{0}(\bar{\alpha}+\bar{\alpha}^{*})$.
In this picture, the optomechanical interaction has changed the form.
It contains the bilinear term $g_{0}(\hat{a}+\hat{a}^{\dagger})(\bar{\beta}\hat{b}_{1}^{\dagger}+\bar{\beta}^{*}\hat{b}_{1})$
that we introduced in the text, in addition to the original term $g_{0}(\hat{a}+\hat{a}^{\dagger})\hat{b}_{1}^{\dagger}\hat{b}_{1}$.
It is clear that the later will be negligible whenever $|\bar{\beta}|^{2}\gg\langle\hat{b}_{1}^{\dagger}\hat{b}_{1}\rangle$.
But there is one more way in which it can become negligible: since
$(\hat{a}+\hat{a}^{\dagger})$ oscillates at frequency $\nu$, the
rotating-wave approximation will suppress it whenever $\nu\gg g_{0}\sqrt{\langle\hat{b}_{1}^{\dagger}\hat{b}_{1}\rangle}$.
Under any of these conditions, the Hamiltonian can then be approximated
by only its bilinear term, as we provided in (5) in the main text. Moreover, working
in the regime $\Delta_{1}=\nu\gg\mathrm{max}\{g_{0}|\bar{\beta}|,\gamma,\Gamma\}$
(required for a proper beam splitter operation) allowed us to made the approximation $\tilde{\Delta}_{1}\approx\Delta_{1}$ in the main text.

\vspace{3mm}
\begin{center}
\textbf{\large{}III. Effect of parameter fluctuations}{\large\par}
\end{center}

Here we explain in detail how we have analyzed the effect of parameter fluctuations in our protocol. The basic idea is that, instead of considering
ideal operations, we consider a beam splitter $\hat{B}_{\lambda}=\exp[\lambda(\hat{a}\hat{b}_{1}^{\dagger}-\hat{a}^{\dagger}\hat{b}_{1})]$ and a controlled-$\theta$ $\hat{C}_{\theta}=\exp(i\theta\hat{b}_{2}^{\dagger}\hat{b}_{2}\hat{a}^{\dagger}\hat{a})$ with fluctuating parameters $\lambda=\pi/4+\delta\lambda$ and $\theta=\pi+\delta\theta$, where both fluctuations $\delta\lambda$ and $\delta\theta$ are taken as Gaussian stochastic processess. Denoting either of them by $\delta z$ (hence $z=\lambda,\theta$), we then have
\begin{equation}
\overline{\delta z^{n}}=\left\{ \begin{array}{cc}
0 & \text{for }n\in\text{odd}\\
(n-1)!!\hspace{1mm}V_{z}^{n/2} & \text{for }n\in\text{even}
\end{array}\right.,\label{StochasticMoments}
\end{equation}
where $V_z$ is the variance (square of the standard deviation)
of the fluctuations, and in the following we denote stochastic averages
by an overbar.

As a proof of principle, we then evaluate the (stochastically-averaged)
fidelity between the ideal and fluctuating states. In order to simplify
the calculation, we will consider the states right after the controlled-$\theta$
operation instead of at the very end of the protocol. In any case,
this will give us a fair idea of the sensitivity of the protocol to
parameter fluctuations. Consider then the state at this stage of the
protocol, which can be written as
\begin{equation}
|\Phi_{\lambda,\theta}\rangle=\frac{1}{\sqrt{2}}\hat{C}_{\theta}\hat{B}_{\lambda}|N\rangle_{1}|0\rangle_{a}(|0\rangle_{2}+|1\rangle_{2})=\frac{1}{\sqrt{2}}\left(|\psi_{\lambda}\rangle_{1a}|0\rangle_{2}+\hat{P}_{\theta}|\psi_{\lambda}\rangle_{1a}|1\rangle_{2}\right),
\end{equation}
where $\hat{P}_{\theta}=e^{i\theta\hat{a}^{\dagger}\hat{a}}$ and
\begin{equation}\label{psi1a}
|\psi_{\lambda}\rangle_{1a}=\hat{B}_{\lambda}|N\rangle_{1}|0\rangle_{a}=\sum_{k=0}^{N}\sqrt{\left(\begin{array}{c}
N\\
k
\end{array}\right)}\sin^{k}\lambda\cos^{N-k}\lambda|N-k\rangle_{1}|k\rangle_{a}.
\end{equation}
The overlap between this state and the ideal one $|\Phi_{\pi/4,\pi}\rangle$
is then given by
\begin{equation}
\langle\Phi_{\pi/4,\pi}|\Phi_{\lambda,\theta}\rangle=\frac{1}{2}\left(\langle\psi_{\pi/4}|\psi_{\lambda}\rangle_{1a}+\langle\psi_{\pi/4}|\hat{P}_{\delta\theta}|\psi_{\lambda}\rangle_{1a}\right),
\end{equation}
where we have used $\hat{P}_{\pi}^{\dagger}\hat{P}_{\theta}=\hat{P}_{\delta\theta}$.
Using (\ref{psi1a}), these two terms are easily rewritten as\begin{subequations}
\begin{align}
\langle\psi_{\pi/4}|\psi_{\lambda}\rangle_{1a} & =\sum_{k=0}^{N}\frac{1}{\sqrt{2^{N}}}\left(\begin{array}{c}
N\\
k
\end{array}\right)\sin^{k}\lambda\cos^{N-k}\lambda,\\
\langle\psi_{\pi/4}|\hat{P}_{\delta\theta}|\psi_{\lambda}\rangle_{1a} & =\sum_{k=0}^{N}\frac{1}{\sqrt{2^{N}}}\left(\begin{array}{c}
N\\
k
\end{array}\right)e^{ik\delta\theta}\sin^{k}\lambda\cos^{N-k}\lambda,
\end{align}
\end{subequations}so that
\begin{equation}
\langle\Phi_{\pi/4,\pi}|\Phi_{\lambda,\theta}\rangle=\sum_{k=0}^{N}\frac{1}{2\sqrt{2^{N}}}\left(\begin{array}{c}
N\\
k
\end{array}\right)\left(1+e^{ik\delta\theta}\right)\sin^{k}\lambda\cos^{N-k}\lambda.
\end{equation}
On the other hand, the average fidelity can be evaluated as
\begin{equation}
\mathcal{F}=\overline{|\langle\Phi_{\pi/4,\pi}|\Phi_{\lambda,\theta}\rangle|},\label{Fidelity}
\end{equation}
that is, the average of the absolute value of the overlap. Let us
then now perform the required stochastic averages. First, let us note
that\begin{subequations}\label{StochasticBasicAverages}
\begin{align}
\overline{e^{inz}} & =e^{-n^{2}V_{z}/2},\\
\overline{\sin^{n}\lambda\cos^{m}\lambda} & =\frac{1}{2^{n+m}i^{n}}\sum_{l=0}^{n}\sum_{l'=0}^{m}(-1)^{l}\left(\begin{array}{c}
n\\
l
\end{array}\right)\left(\begin{array}{c}
m\\
l'
\end{array}\right)e^{i\pi(n+m-2l-2l')/4}e^{-(n+m-2l-2l')^{2}V_{\lambda}/2},
\end{align}
\end{subequations}expressions that we prove at the end of the section.
Hence, we can write
\begin{align}
\overline{\langle\Phi_{\pi/4,\pi}|\Phi_{\lambda,\theta}\rangle} & =\sum_{k=0}^{N}\frac{1}{2\sqrt{2^{N}}}\left(\begin{array}{c}
N\\
k
\end{array}\right)\left(1+\overline{e^{ik\delta\theta}}\right)\overline{\sin^{k}\lambda\cos^{N-k}\lambda}\\
 & \hspace{-1cm}=\sum_{k=0}^{N}\frac{1}{2\sqrt{2^{N}}}\left(\begin{array}{c}
N\\
k
\end{array}\right)\left(1+e^{-k^{2}V_{\theta}/2}\right)\frac{1}{2^{N}i^{k}}\sum_{l=0}^{k}\sum_{l'=0}^{N-k}(-1)^{l}\left(\begin{array}{c}
k\\
l
\end{array}\right)\left(\begin{array}{c}
N-k\\
l'
\end{array}\right)e^{i\pi(N-2l-2l')/4}e^{-(N-2l-2l')^{2}V_{\lambda}/2}\nonumber \\
 & =\sum_{k=0}^{N}\sum_{l=0}^{k}\sum_{l'=0}^{N-k}\frac{(-1)^{l}}{2^{1+3N/2}i^{k}}\left(\begin{array}{c}
N\\
k
\end{array}\right)\left(\begin{array}{c}
k\\
l
\end{array}\right)\left(\begin{array}{c}
N-k\\
l'
\end{array}\right)e^{i\pi(N-2l-2l')/4}\left(1+e^{-k^{2}V_{\theta}/2}\right)e^{-(N-2l-2l')^{2}V_{\lambda}/2},\nonumber 
\end{align}
expression that can be evaluated very efficiently in any computer.

\begin{figure}[t]
\centering
\includegraphics[width=0.9\textwidth]{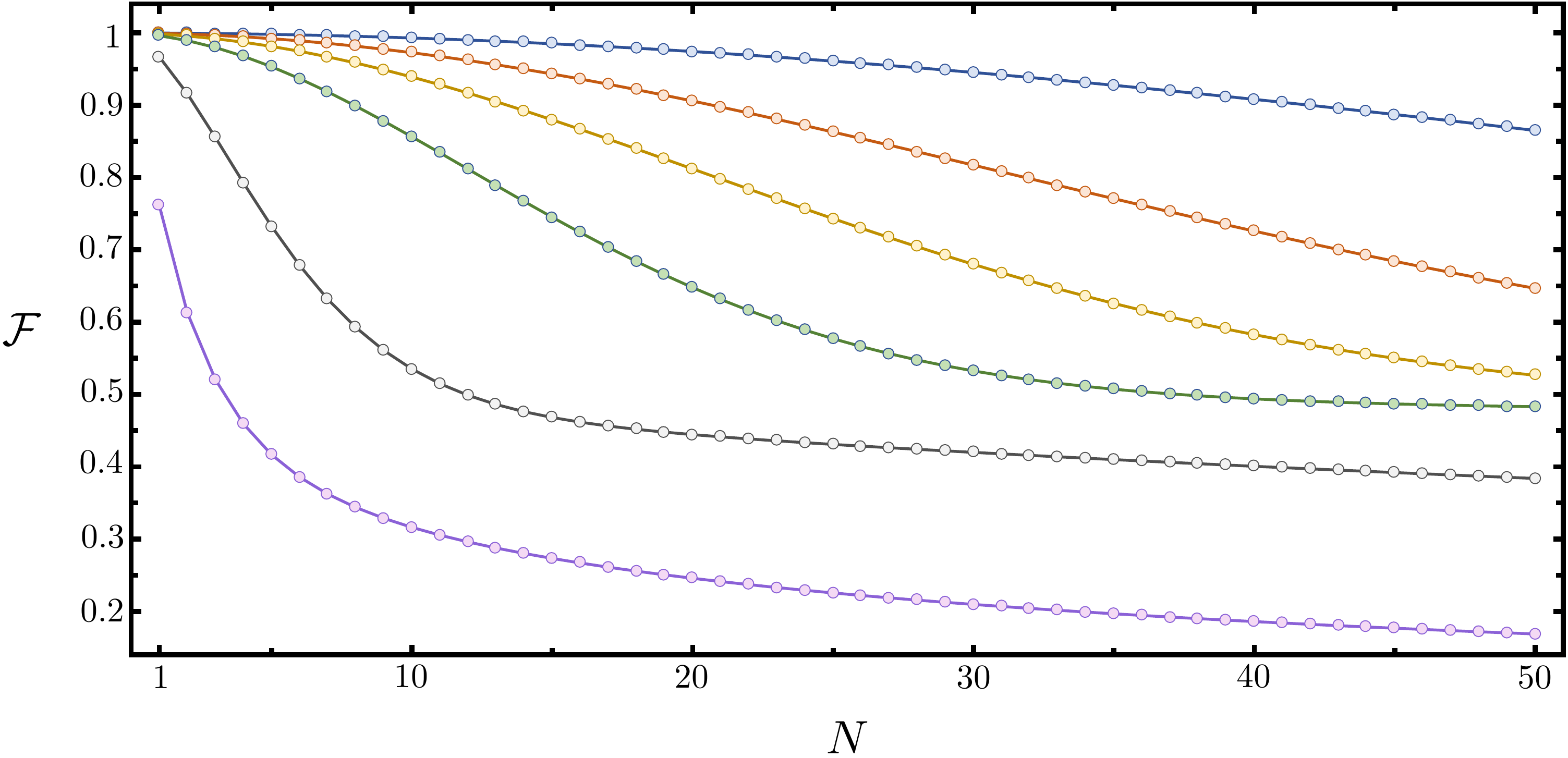}
\caption{Fidelity of the state generated by our protocol as a function of the N00N size $N$, and for different values of the standard deviation of the beam splitter and controlled-$\pi$ parameters: 1\% (blue), 2\% (orange), 3\% (yellow), 5\% (green), 15\% (grey), and 50\% (purple), from top to bottom. \label{Fig:Fidelity}}
\end{figure}

In Fig. \ref{Fig:Fidelity} we plot the fidelity (\ref{Fidelity}) as a function of
$N$ for different values of the standard deviation of the parameters.
As mentioned in the main text, for 1\% standard deviation, the fidelity
stays above 90\% for values as large as $N=40$. Note that the standard deviation is the square root of the variance, and hence, $M$\% means $V_\theta=(10^{-2}M\pi)^2$ and $V_\lambda=(10^{-2}M\pi/4)^2$.

Let us now prove expressions (\ref{StochasticBasicAverages}). In
the case of the first one, we simply expand the exponential in tailor
series and use (\ref{StochasticMoments}), leading to
\begin{align}
\overline{e^{inz}} & =\sum_{k=0}^{\infty}\frac{1}{k!}(in)^{k}\overline{z^{k}}=\sum_{l=0}^{\infty}\frac{(2l-1)!!}{(2l)!}(in)^{2l}V_{z}^{l}=\sum_{l=0}^{\infty}\frac{1}{2^{l}l!}(-n^{2})^l V_{z}^{l}=e^{-n^{2}V_{z}/2}.
\end{align}
As for the second expression, it is also easy to prove by writing
the trigonometric functions in terms of complex exponentials and using
the previous expression: 
\begin{align}
\overline{\sin^{n}\lambda\cos^{m}\lambda} & =\frac{1}{2^{n+m}i^{n}}\overline{\left(e^{i\lambda}-e^{-i\lambda}\right)^{n}\left(e^{i\lambda}+e^{-i\lambda}\right)^{m}}=\frac{1}{2^{n+m}i^{n}}\sum_{l=0}^{n}\sum_{l'=0}^{m}(-1)^{l}\left(\begin{array}{c}
n\\
l
\end{array}\right)\left(\begin{array}{c}
m\\
l'
\end{array}\right)\overline{e^{i(n+m-2l-2l')\lambda}}\\
 & =\frac{1}{2^{n+m}i^{n}}\sum_{l=0}^{n}\sum_{l'=0}^{m}(-1)^{l}\left(\begin{array}{c}
n\\
l
\end{array}\right)\left(\begin{array}{c}
m\\
l'
\end{array}\right)e^{i\pi(n+m-2l-2l')/4}e^{-(n+m-2l-2l')^{2}V_{\lambda}/2}.\nonumber 
\end{align}

\newpage
\end{widetext}

\end{document}